# Beam Shifting due to Bifurcation in a Cavity Environment

Carol Scarlett
Physics Department, Florida A&M University

Abstract:
A number of physical processes show some form of bifurcation or periodic splintering of single distributions into two new ones. Recently, it has been noted that cavity searches for interactions between photons and exotic fields may also result in bifurcation[1]. This paper builds on previous simulations of bifurcation of an optical beam in the presence of periodic focusing[2]. Here, however, the focus is on predicting a shifting of the beam's position, defined by the center of the energy density that can result. Mathematical models are described and the formalism for simulating bifurcation under complex conditions is delineated.

**I. Introduction:**

Bifurcation is a well explored mathematical phenomena with applications in Chaos Theory, Turbulence[3] and studies of population growth[4-5]. Recently, preliminary work shows that the pseudo particle states that can be formed from photon and axion field interactions will split when exposed to an external, inhomogeneous magnetic field. These states can be collapsed through interactions with a mirror surface leading to a splitting of the photon distributions with each reflection. Conventionally, optical cavities utilize curved mirrors that periodically focus a beam. The effect of periodic focusing was explored in an earlier work (see ref. [2] on measurement of bifurcation). The current work focuses on shifting of the energy distribution when the angle of injection into a cavity is nonzero. The shifting occurs as a result of the geometry of the cavity and the enhancement of transverse momentum for each new photon distribution. A full simulation of bifurcation with focusing and non-zero angle of injection into a cavity environment is described below.

In Section II a Taylor expansion of the function describing the beam position reveals how a shifting of the energy density arises, when higher order terms are considered. The approach taken involves expansion of the deflection angle to calculate the impact of shifting on measurable parameters. Section III develops the basic algebra when considering transition of a photon beam through a cavity. The ABCD Matrices are described along with the geometry used throughout the remainder of the paper. Section IV discusses the outcome of simulations for various cavity types. Graphs depict the impact of focusing on the magnitude of the energy shifting. The Conclusion section explores the application of these results to cavity searches for axions and axion like particles.

**II. Higher Order Terms**

The process of bifurcation of a photon beam due to induced interactions between the beam and exotic particles such as axions has been investigated in both ref. [1] and [2]. There it was shown that as a beam entered a region of inhomogeneous magnetic field, it would split resulting in two new distributions propagating with an angle $+ 2 \cdot \theta$ relative to one another. The notation previously used referred to the direction and angle for these new distributions as a "Chief Ray." Ref. [2] explored, for a Fabre-Perot cavity environment, the nature of periodic splitting along with focusing and propagating. There an expression was derived for a potentially measurable result. If, however, the cavity is an optical delay the beam can be injected at a non-zero initial angle. In this case, the splitting angle can couple to the angle of injection such that a net shift of the energy distribution results. Thus bifurcation leads to a shifting of the energy distribution. When the splintered distributions overlap significantly, as for small splitting angles, this effect is measured as a shifting of the center of the beam or deflection of the entire distribution.

To understand the origins and amplitude of beam shifting, one can start with a Taylor Expansion of the angular function $\tan(\theta)$ used to calculate the beam position:

$$\tan(\theta) = \theta + \frac{1}{3}\theta^3 + \frac{2}{15}\theta^5 + \frac{17}{315}\theta^7 \ldots \tag{1}$$

If two rays emerge from a parent ray making angles $\pm\theta_{split}$ and the parent ray makes an angle $\theta_i$ relative to a central axis, then the appropriate angles for the new rays can be written as $\theta_i \pm \theta_{split}$. Replacing the angle in the tangent function (1) and keeping the $\theta^3$ term gives the next higher order correction to the position of each new distribution. Below, this term for the two new chief rays (D1 and D2) is compared to the same term for the parent ray to assess the change in the position of the energy center for the outgoing distribution:

$$ParentRay: (\theta_i)^3 = \theta_i^3$$
$$D1: \frac{1}{2}(\theta_i + \theta_{split})^3 = \frac{1}{2}[\theta_i^3 + 3\theta_i\theta_{split}^2 + 3\theta_i^2\theta_{split} + \theta_{split}^3]$$
$$D2: \frac{1}{2}(\theta_i - \theta_{split})^3 = \frac{1}{2}[\theta_i^3 + 3\theta_i\theta_{split}^2 - 3\theta_i^2\theta_{split} - \theta_{split}^3]$$
$$\overline{D1 + D2 = \theta_i^3 + 3\theta_i\theta_{split}^2} \tag{2}$$

Even for a single pass through the cavity the weighted position, i.e. multiplying each chief ray by half the intensity of the parent ray and summing, includes a term $3\theta_i \cdot \theta_{split}^2$ that adds positively instead of cancelling. This means that the new rays have an average position shifted by this amount relative to the Parent Ray. For splitting angles consistent with axion theory, of order $\theta_{split}^2 = (10^{-15})^2$, this appears infinitesimal. However, in a cavity the effect can build quite rapidly.

To fully appreciate this shifting effect, consider what happens to the density profile for a beam with an non-zero injection angle into an optical delay. Starting with a Gaussian function that describes the distribution of photons $P_D$:

$$P_D = Ae^{-\frac{1}{2}[\frac{x}{r}]^2} \tag{3}$$

Where x represents position starting from the center of the distribution, implicitly zero in the expression above, r defines the beam waist, and $2 \cdot r$ represents the $e^{-2}$ position where the beam is ~ 1/7 of its maximum. For the case of a single pass through an optical cavity, if both splitting and shifting occurs due to coupling of the beam to an axion field, the expression above can be rewritten in terms of two new distributions that result:

$$P_D' = A \cdot \frac{1}{2} \cdot e^{-\frac{1}{2}[\frac{x-\alpha+\eta}{r}]^2}$$

$$P_D'' = A \cdot \frac{1}{2} \cdot e^{-\frac{1}{2}[\frac{x+\alpha+\eta}{r}]^2} \qquad (4)$$

Where $\alpha$ represents a displacement of the new distribution's center, in a positive direction, due only to the induced splitting and $\eta$ represents the additional displacement due to the higher order coupling between the injection angle and splitting. The splitting leads to a movement of energy away from the center of the distribution into the sidebands. The shifting effect, on the other hand, is characterized by a movement of all energy either towards the left or the right of the unaltered distribution. For this reason it is convenient to rewrite (4) in terms of the energy density left $P_L$ and right $P_R$. This yields:

$$P_L = A \cdot \frac{1}{2} \cdot e^{-\frac{1}{2}[\frac{x}{r}(1-\frac{\alpha}{x}+\frac{\eta}{x})]^2} \qquad x < 0$$

$$P_R = A \cdot \frac{1}{2} \cdot e^{-\frac{1}{2}[\frac{x}{r}(1+\frac{\alpha}{x}+\frac{\eta}{x})]^2} \qquad x > 0 \qquad (5)$$

For the theoretical values of splitting the measurables scale as: $\eta \sim 10^{-30}$ and $\alpha \sim 10^{-15}$. Thus we keep only terms involving the first power in $\eta$ and up to $\alpha^2$, neglecting both $\eta^2$ and $\alpha \bullet \eta$ terms. Squaring the expression in brackets yields:

$$(1-\frac{\alpha}{x}+\frac{\eta}{x})^2 = 1 - \frac{2\alpha}{x} + \frac{2\eta}{x} + \frac{\alpha^2}{x^2} - \frac{2\alpha\eta}{x^2} + \frac{\eta^2}{x^2}$$

$$P_L \approx A \cdot \frac{1}{2} \cdot e^{-\frac{1}{2}[\frac{x}{r}(1-\frac{\alpha}{x}+\frac{\eta}{x})]^2} \approx \frac{A}{2} \cdot e^{-[\frac{x^2}{r^2}(\frac{1}{2}-\frac{\alpha}{x}+\frac{\alpha^2}{2x^2}+\frac{\eta}{x}+...)]} \qquad x < 0 \qquad (6)$$

$$P_R \approx A \cdot \frac{1}{2} \cdot e^{-\frac{1}{2}[\frac{x}{r}(1+\frac{\alpha}{x}+\frac{\eta}{x})]^2} \approx \frac{A}{2} \cdot e^{-[\frac{x^2}{r^2}(\frac{1}{2}+\frac{\alpha}{x}+\frac{\alpha^2}{2x^2}+\frac{\eta}{x}+...)]} \qquad x > 0$$

Subtracting $P_R$ from $P_L$ gives the degree of shifting for the photon density as compared to the case of no splitting.

$$P_L - P_R \approx A \cdot \frac{1}{2} \cdot e^{-\frac{1}{2}\frac{x^2}{r^2}} \cdot e^{-\frac{\alpha^2}{2r^2}} \cdot e^{\frac{\alpha \cdot x}{r^2}} \cdot \left\{ e^{\frac{-\eta \cdot x}{r^2}} - e^{\frac{\eta \cdot x}{r^2}} \right\}$$

$$P_L - P_R \approx Ae^{-\frac{1}{2}\frac{x^2}{r^2}}[(1-\frac{\alpha^2}{2r^2}) \cdot \sinh(\frac{\eta \cdot x}{r^2})] \qquad (7)$$

$$P_L - P_R \approx Ae^{-\frac{1}{2}\frac{x^2}{r^2}}[\frac{\eta \cdot x}{r^2}]$$

Finally, one derives an expression directly proportional to the shifting that results from a non-zero injection (proportional to the square of $g_a$). The final form rises as a sinh function with $\eta$.

In a cavity environment, the process of focusing, propagating and re-splitting makes both α and η complicated functions of the number of transitions through the cavity. As in Ref. [2], α can be expressed in terms of the splitting angle ($\theta_{split}$), cavity length (d), and a function $f_{split}(n)$ that measures the effective movement of the energy away from the center and into sidebands:

$$\alpha = \theta_{split} \cdot d \cdot f_{split}(n) \tag{8}$$

The function $f_{split}(n)$ must be extracted from simulations of the photon density where each new chief ray is taken into account. Likewise, the shifting variable η can also be expressed as:

$$\eta = 3 \cdot \theta_{injection} \cdot \theta_{split}^2 \cdot d \cdot f_{injection}(n) \tag{9}$$

As with the case for splitting, the $f_{injection}(n)$ function characterizes how a signal, due to non-zero injection coupled with splitting, builds in a cavity and accounts for real effects such as focusing.

The next section derives the formulism to describe both splitting and shifting. In what follows various cavity scenarios are explored and both $f_{split}(n)$ and $f_{injection}(n)$ functions are extracted.

**III. Beam Dynamics:**

In optics, paraxial photon beams are often described by a one-dimensional vector of the form:

$$\begin{bmatrix} X_0 \\ \theta_i \end{bmatrix} \tag{10}$$

where $X_0$ represents the initial position of a ray as measured from a central axis and $\theta_i$ represents the incoming angle that the ray makes relative to some specified path. Figure 1 shows the geometry for this initial vector.

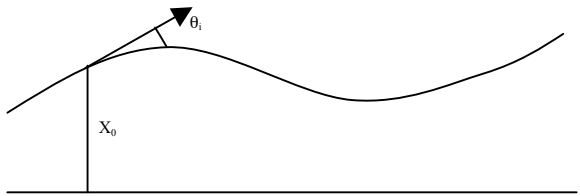

Figure 1: An input vector to an optical system.

For a beam, we can take $X_0$ to be the radius at $e^{-1}$ and $\theta_i$ to be the divergence.

To transform some initial vector into a new one, following focusing or propagation in a cavity, we use the ABCD Matrix formulism. These matrices are written as:

$$\text{Focusing:} \quad \begin{bmatrix} 1 & 0 \\ \dfrac{-1}{f} & 1 \end{bmatrix} \quad (11)$$

where f represents the focal length of a lens.

$$\text{Propagation:} \quad \begin{bmatrix} 1 & d \\ 0 & 1 \end{bmatrix} \quad (12)$$

where d represents the distance traveled.

The same formalism can be used to describe a process whereby one ray splits into two new rays. Equation (13) shows a matrix that takes a single ray and transforms it into two new rays having angles $+\theta_{split}$ and $-\theta_{split}$ relative to some initial, "incoming", angle $\theta_i$:

$$\text{Splitting:} \quad \begin{bmatrix} 1 & 0 \\ \dfrac{\pm \theta_{split}}{X_0} & 1 \end{bmatrix} \quad (13)$$

Conventionally, these matrices are applied to a vector, such as in (10), that describes the incoming radius and waist of a Gaussian beam. Here, in addition to using the formulism to determine the resulting vectors, the matrices are used to track the chief rays, which give the position and direction for each distribution as it emerges and propagates through the cavity.

Figure 2 shows a typical cavity setup for experiments searching for exotic particles such as axions. As discussed in ref. [2], by summing over all final distributions, the changes in the beam profile are calculated. Equation (14) gives the single pass application of the formulism:

$$\begin{bmatrix} 1 & (E-D) \\ 0 & 1 \end{bmatrix} \begin{bmatrix} 1 & (D-C) \\ 0 & 1 \end{bmatrix} \begin{bmatrix} 1 & 0 \\ \dfrac{(\pm)\theta_{split}}{R^*} & 1 \end{bmatrix} \begin{bmatrix} 1 & (C-B) \\ 0 & 1 \end{bmatrix} \begin{bmatrix} 1 & 0 \\ \dfrac{\pm \theta_{split}}{R^*} & 1 \end{bmatrix} \begin{bmatrix} 1 & (B-A) \\ 0 & 1 \end{bmatrix} \begin{bmatrix} R_0 \\ \theta_{injection} \end{bmatrix} \quad (14)$$

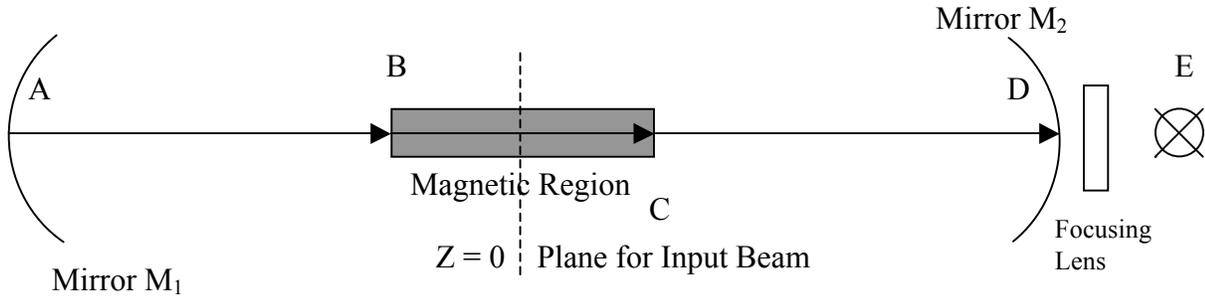

Figure 2: Sketch for a possible experimental setup showing a mirror cavity (A to D), magnetic field region (B to C) and detector (E).

Applying the algebra in (14) yields equation (15). For multiple traversals of the cavity, one arrives at equation (16):

$$\begin{bmatrix} R_1 \\ \theta_1 \end{bmatrix} = \begin{bmatrix} 1 \pm \dfrac{\theta_{split} \cdot (d + 2 \cdot |E - D|)}{R_0} & d + |E - D| \\ \dfrac{\pm 2 \cdot \theta_{split}}{R_0} & 1 \end{bmatrix} \begin{bmatrix} R_0 \\ \theta_{injection} \end{bmatrix} \qquad (15)$$

$$\begin{bmatrix} R_n \\ \theta_n \end{bmatrix} = \begin{bmatrix} [1 - \dfrac{d}{f} \pm \dfrac{\theta_{split} \cdot d}{R_1}] & d \\ \dfrac{-1}{f} \pm \dfrac{2 \cdot \theta_{split}}{R_1} & 1 \end{bmatrix}^{n-1} \cdot \begin{bmatrix} R_1 \\ \theta_1 \end{bmatrix} \qquad (16)$$

Note, after each application of (15) the number of subsequent distributions increases by a factor of 2. Thus after n application there are $2^n$ distributions that must be weighted and summed. These two expressions show the roles of mirror focusing (f), cavity length (d), and input angle ($\theta_{injection}$) on the position of each new distribution.

Table 1 gives some typical experimental parameters starting around the current limit for traditional cavity searches ($g_a \sim 10^{-6}$).

Table 1:

| Cavity Type | Cavity Length | Magnetic Field Length | VB Strength | Laser Wavelength | Laser Energy | Mirror Radius | Number of Bounces | $\theta_{split} \sim 10^{-10}$ ($g_a = 10^{-6}$) | Injection Angle |
|---|---|---|---|---|---|---|---|---|---|
| Confocal | 14 m | 10 m | 200 T/m | 1064 nm | 1 W | 25 m | $1.2 \cdot 10^4$ | $4 \cdot 10^{-10}$ | $1.5 \cdot 10^{-2}$ rad |
| Convex-Concaved | 14 m | 10 m | 200 T/m | 1064 nm | 1 W | 25 m, -11 m | $1.2 \cdot 10^4$ | $4 \cdot 10^{-10}$ | $1.5 \cdot 10^{-2}$ rad |
| Planear-Planear | 14 m | 10 m | 200 T/m | 1064 nm | 1 W | $\infty$ | $1.2 \cdot 10^4$ | $4 \cdot 10^{-10}$ | $1.5 \cdot 10^{-2}$ rad |

The magnitude of the effect following a single pass can be calculated using equation (7). Using the parameters in Table 1 and equation (7), the effect can be estimated:

$$P_L - P_R \approx A e^{-\frac{1}{2} \frac{x^2}{r^2}} [(1 - \frac{\alpha^2}{2r^2}) \sinh(\frac{\eta \cdot x}{r^2})] \approx A e^{-\frac{1}{2} \frac{x^2}{r^2}} [\frac{x \cdot \eta}{r^2}]$$

*Considering only the changes in density due to the first term and taking $x \sim r$*

*The integrated difference can be approximated as a triangle of base r and height $(P_L - P_R)\big|_{x \sim r}$* (17)

$$\int (P_L - P_R) \cdot dr \approx \frac{5}{6} \cdot 10^{+18} \cdot 0.61 \cdot [\frac{3 \cdot (4 \cdot 10^{-10})^2 \cdot 1.5 \cdot 10^{-3} \cdot 10}{7.5 \cdot 10^{-4}}] \approx 4.86 \ photons/\sec$$

*Where A has been normalized over the interval.*

The observable is a drop in the intensity measured to the left (right) of the beam center for the zero-splitting case accompanying an increase in the intensity as measured to the right (left) of the same center. It is well known that the center of any laser beam is not infinitely stable. However, pointing instability leads to "white" or random noise. This means that the beam shifts randomly regardless of whether or not there is a splitting of the beam. For this reason, modulation of the effect, accomplished by modulation of the beam, must be used to distinguish a signal from the background noise.

Clearly, the loss indicated in equation (17) is well below the shot noise value of $\approx 9.12 \cdot 10^8$ photons/sec. However, as seen for the case of bifurcation in a cavity environment, these effects often grow rapidly enough that they become measurable after some number of traversals through the cavity. The next section explores the pattern of growth expected for each traversal of the cavity and shows the predictions from simulations for the three cavity types in Table 1.

## IV. Simulations

It is worthwhile to discuss how a left-right energy difference can be experimentally measured and why this effect can be separately consider in addition to the spreading of energy away from the center. Importantly, the shifting measurement is made as a comparison between the location of the energy center for the case where the magnetic field gradient is zero – no splitting occurs and therefore the η function is explicitly zero – and the case of maximal field. For a modulated external magnetic field gradient, the shifting becomes a signal at the modulation frequency. This signal can be recorded on a quadrant photo-detector with left-right and up-down divisions. The signal itself is simply a shifting of the beam position recorded as a change in the energy difference between halves of the detector. Because this signal does not involve, or to first-order couple with, movement of photonic energy away from the peak it is distinct from the signal due to splitting (represented by the function α).

As many resonator cavities are constructed with confocal mirrors, this is the first experimental scenario considered. Simulation of the splitting and shifting of the beam have been carried out, using the parameters in Table 1, to determine the function $f_{injection}(n)$. Figure 3 shows the degree of shifting as a function of traversals of the cavity by the beam.

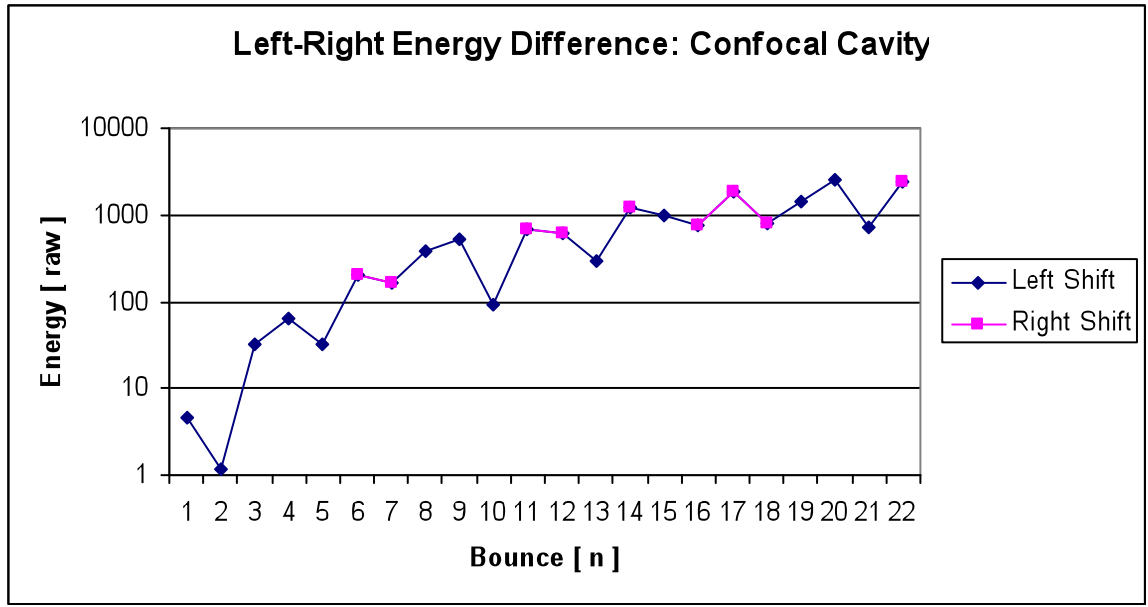

Figure 3: Shifting of the beam center, relative to the case of $\theta_{split} = 0$, to either the left (blue) or right (pink) with each passage through the cavity.

Here it can be seen that the magnitude of the shifting rises quite rapidly. After only 20 traversals of the cavity, the energy difference has risen by a factor ~2000. While there are fluctuations, the magnitude consistently and sharply increases. Several fits were made to extract the $f_{injection}(n)$. The best $\chi^2$ achieved with standard fit options (linear, polynomial, exponential, etc.) was ~ 0.86 and gave a function of $1.8*x^{2.27}$ that appears in figure 4 below (blue). The data also appears in the figure for comparison.

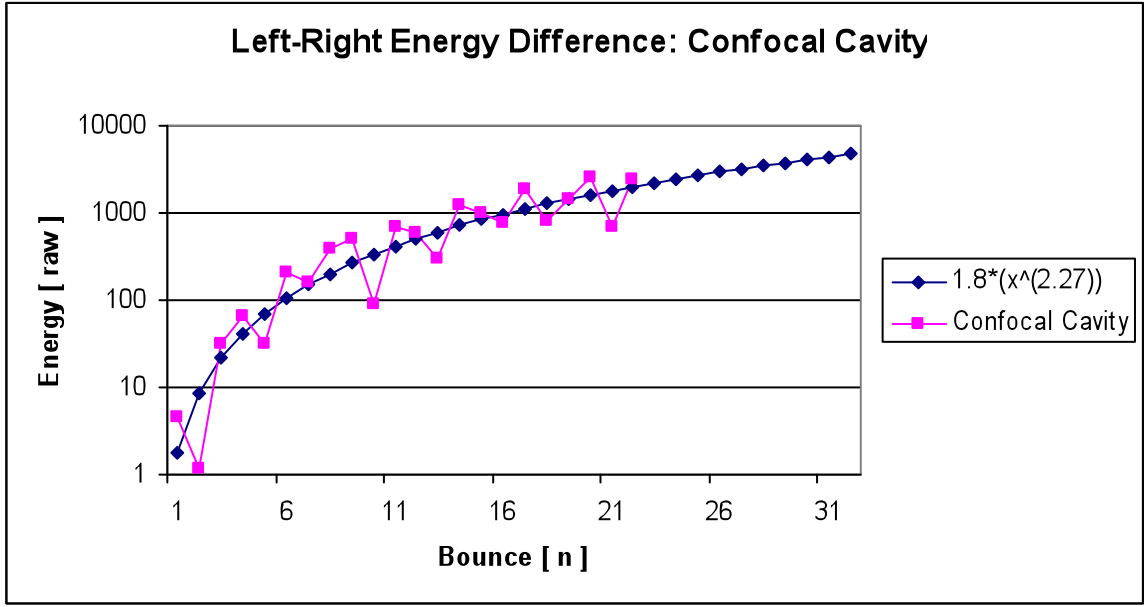

Figure 4: Simulation of the left-right energy difference (pink) plotted against a fit function (blue) for comparison.

The fit can now be used to estimate the size of a signal following any number of transitions through a confocal cavity with the conditions defined in the Table 1. As an example, equation (18) gives the energy difference between the left and right halves:

$$Left-Right\ \ Energy\ \ Difference:$$
$$1.8 \cdot (12000\,transits)^{2.27}\,\frac{photons}{transit} \approx 3.27E+09\ \ photons/\sec \quad (18)$$

When compared to the shot noise level, ~ $9.12 \cdot 10^{+08}$ photons/sec, the shifting for the cavity described above appears measurable. In addition to calculating shifting, the position of the beam in the delay was also tracked. Not surprising, for the injection angle, the position does not change considerably indicating that the beam does not wonder outside a cavity with ~ 4 cm diameter.

As with the case of splitting, the shifting depends on cavity parameters such as focusing and length. Thus, the type of cavity must be considered. Figure 5 (panel a) shows the simulation of shifting in a concave-plane cavity. The growth of shifting with transits through the cavity look much like those for the confocal cavity; in fact the fit functions are approximately the same as seen in panel b.

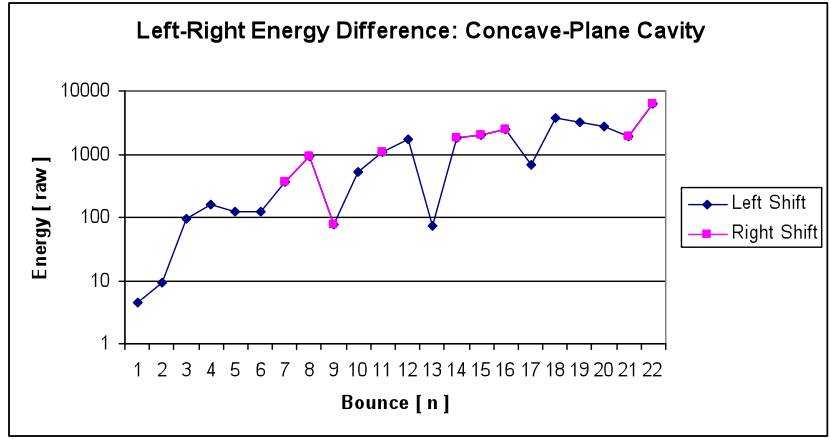

(a)

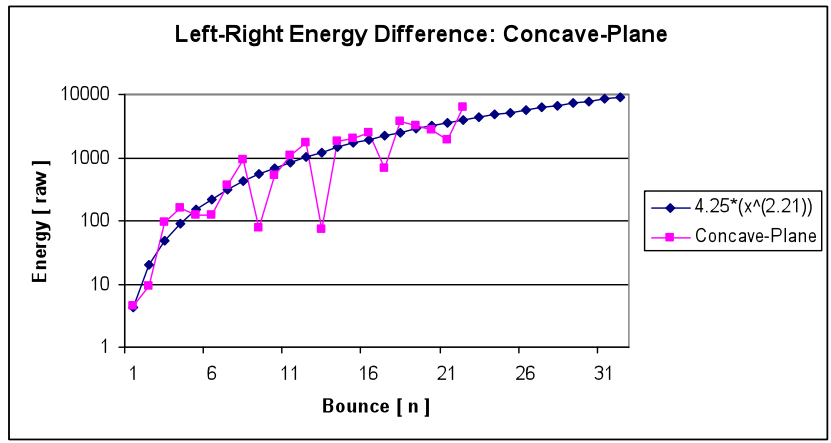

(b)

Figure 5: (a) Shifting of the beam center, relative to the case of $\theta_{split} = 0$, to either the left (blue) or right (pink) with each passage through the cavity; (b) Simulation of the left-right energy difference (pink) plotted against a fit function (blue) for comparison.

The extracted η function does not deviate much from the confocal case. Qualitatively, the beam appears to shift back and forth about the center with values that drop closer to zero when compared to the confocal case. This leads to a somewhat worse fit, $\chi^2 \sim 0.7987$. The final predictions for a measurable change are calculated in equation (19):

$$Left-Right \quad Energy \quad Difference:$$
$$4.24 \cdot (12000 \, transits)^{2.21} \frac{photons}{transit} \approx 4.39E+09 \quad photons/\sec \quad (19)$$

suggest little quantitative change for this cavity compared to the confocal case.

Finally, the concave-convex cavity simulation appears in Figure 6 panel (a) along with a fit function for comparison in panel (b). It is easy to see that the η function rises significantly faster than either the confocal or concave-plane cases.

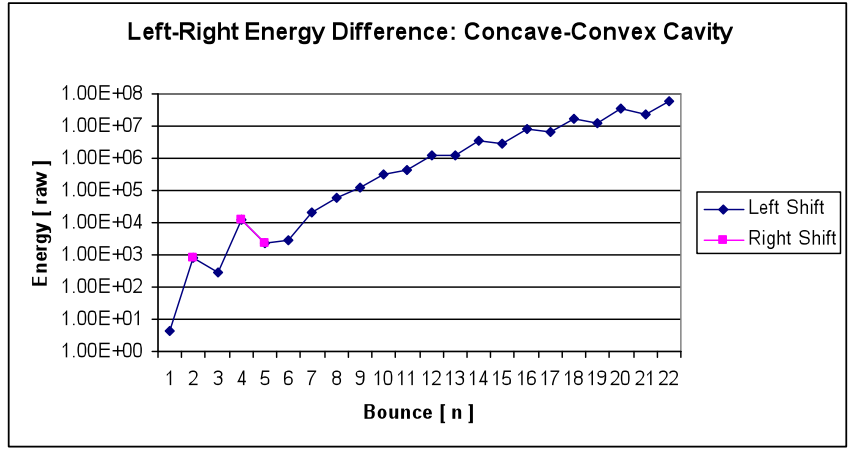

(a)

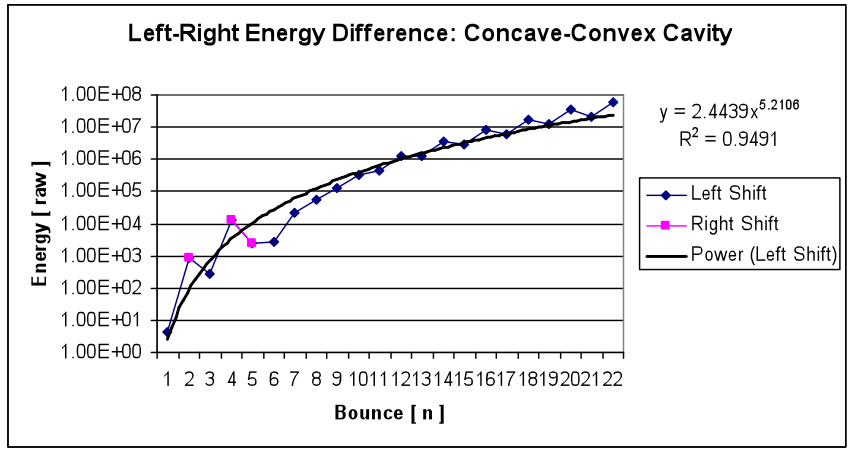

(b)

Figure 6: (a) Shifting of the beam center, relative to the case of θ$_{split}$ = 0, to either the left (blue) or right (pink) with each passage through the cavity; (b) Simulation of the left-right energy difference (pink) plotted against a fit function (blue) for comparison.

If a signal could be extracted after some $1.2 \cdot 10^{+04}$ bounces,

$$Left-Right \quad Energy \quad Difference:$$
$$2.44 \cdot (12000 transits)^{5.21} \frac{photons}{transit} \approx 4.36E+21 \quad photons/\sec \tag{20}$$

NOTE: The final numbers is greater than the total photons impinging on the detector, reflecting the fact that the shifting would cause the beam center to more a distance that is greater than its diameter; completely depleting half of the detector of photons. However, the shifting, see Table 1, requires a coupling $g_a \sim 10^{-06}$ which is significantly larger than the known limit of $\sim 10^{-10}$. Furthermore, the effect scales as $g_a^2$, therefore at the current limit the maximal change would be closer to $\sim 4 \cdot 10^{+13}$. This is significantly larger than the shot noise. FINAL NOTE: While the beam in either a confocal or concave-plane cavity continually reflect towards the cavity center, the concave-convex optical delay shows a rapid departure of the beam center from the cavity center (~ 0.47 m after only 10 transitions) making

this scenario, as simulated, physically unrealizable. The simulations suggest that this case requires a cavity with significant transverse diameter. Other possibilities (one deformable mirror, parabolic mirrors, etc.) are being investigated.

**V. Conclusion**

Preliminary simulations suggest that there is a shifting effect in addition to a broadening of a photon beam due to photon-axion interactions in an external, inhomogeneous magnetic field. As new experiments are being designed, it is important to understand all measurables, if only to exclude sources of backgrounds. Furthermore, these measurables should be studied in the context of previous searches to understand their limitations as well as predict their outcome. The ability to evaluate higher numbers of bounces, are necessary to improve predictions. Estimate of this shifting may reveal the limits on coupling constants that can be probed by this effect alone.

**VI. Acknowledgements**

This work was made possible in part through a grant from the National Science Foundation (NSF) Center of Research Excellence in Science and Technology (CREST) program. A special thanks goes to Mikhail Khankhasayev for his efforts to help review the final text.

---